\def\year{2021}\relax
\documentclass[letterpaper]{article} 
\usepackage{aaai21}  
\usepackage{times}  
\usepackage{helvet} 
\usepackage{courier}  
\usepackage[hyphens]{url}  
\usepackage{graphicx} 
\usepackage{natbib}  
\usepackage{caption} 
\usepackage{wrapfig}
\usepackage{multirow}
\usepackage{makecell}
\usepackage[symbol]{footmisc}
\title{Characterizing Discourse about COVID-19 Vaccines: A Reddit Version of the Pandemic Story}

\author {
    Wei Wu,
    Hanjia Lyu,
    Jiebo Luo \\
}
\affiliations {
    University of Rochester \\
    wwu22@u.rochester.edu,
    hlyu5@ur.rochester.edu,
    jluo@cs.rochester.edu
}

\begin{document}

\maketitle

\begin{abstract}
It has been one year since the outbreak of the COVID-19 pandemic. The good news is that vaccines developed by several manufacturers are being actively distributed worldwide. However, as more and more vaccines become available to the public, various concerns related to vaccines become the primary barriers that may hinder the public from getting vaccinated. Considering the complexities of these concerns and their potential hazards, this study aims to offer a clear understanding about different population groups’ underlying concerns when they talk about COVID-19 vaccines - particular those active on Reddit. The goal is achieved by applying LDA and LIWC to characterizing the pertaining discourse with insights generated through a combination of quantitative and qualitative comparisons. Findings include: 1) during the pandemic, the proportion of Reddit comments predominated by conspiracy theories outweighed that of any other topics; 2) each subreddit has its own user bases, so information posted in one subreddit may not reach those from other subreddits; 3) since users’ concerns vary across time and subreddits, communication strategies must be adjusted according to specific needs. The results of this study manifest challenges as well as opportunities in the process of designing effective communication and immunization programs.

\end{abstract}

\section{Introduction}
Severe acute respiratory syndrome coronavirus 2 (SARS-CoV-2), known as COVID-19, has hit 222 countries and caused over 2 million deaths in the worldwide as of January 14, 2021\footnote{https://www.worldometers.info/coronavirus/}. Since the World Health Organization (WHO) declared COVID-19 a pandemic\footnote{https://www.who.int/director-general/speeches/detail/who-director-general-s-opening-remarks-at-the-media-briefing-on-covid-19---11-march-2020}, it has caused catastrophic damages to various aspects of the human society, from economy~\cite{fernandes2020economic} to psychology~\cite{atalan2020lockdown}. To minimize the deadly impacts of the pandemic, scientists from different countries have been actively involved in the investigations of effective treatments and vaccines. Staring in January 2021, the vaccines developed by the corporations led by Pfizer and Moderna are being distributed across the U.S., while the Johnson \& Johnson vaccine is also on the way\footnote{https://www.cdc.gov/coronavirus/2019-ncov/vaccines/8-things.html}. Despite the arduous efforts of scientists that have demonstrated the efficacy of possible COVID-19 vaccines and their protection mechanisms to human bodies~\cite{Lipsitch763}, the public attitudes towards COVID-19 vaccines are nothing if not disparate.

As early as of June 2020, a survey covering 13,426 participants in 19 countries already showed that the potential acceptance of a COVID-19 vaccine varies largely from country to country: in China, the acceptance rate of a potentially safe and effective COVID-19 vaccine is 90\%, which is the highest among all countries, while that rate in Russia is only 55\%~\cite{lazarus2020global}. The study suggested that the variation in the vaccine acceptance rates may be a result of the difference in levels of trust in central governments. Another study~\cite{REITER20206500}, which mainly focused on adults in the United States, conducted an online survey with 2,006 qualified participants and investigated factors that influence their acceptability of a COVID-19 vaccine. One of the main implications of this study is that participants whose political beliefs are moderate or liberal have a relatively higher vaccine acceptance rate in this case. There are also many other emerging studies that conducted similar surveys to analyze the acceptability of a COVID-19 vaccine among different social, ethnic groups (e.g., \citet{GOLDMAN20207668}), providing policy makers with a clearer picture of the public’s attitudes towards COVID-19 vaccines. These studies are important because they shed light on the hidden obstacles and challenges in the vaccine distribution process, and offer critical information for implementing appropriate communication strategies in times of crisis. Nevertheless, because of the limitations of the survey approach — including but not limited to a small sample size, a definite number of available choices, etc. — it is difficult to know what the specific concerns of the general public are when they think of COVID-19 vaccinations. In addition, out of the fear of being considered as ``anti-vaxxers", people with concerns may choose to restrain themselves from expressing their genuine opinions in the aforementioned surveys~\cite{brenner2016lies} or not to participate in them at all. 

To supplement and verify the findings in the preceding studies, we employ a data-mining approach with Natural Language Processing (NLP) techniques to investigating the discourse related to COVID-19 vaccines using large-scale social media data. Since the advents of social media platforms, they have been widely used as reliable data sources for a variety of research purposes. From presidential election analyses~\cite{jin2017detection} to electronic cigarette perception examinations~\cite{lu2020user}, social media data demonstrates advantages such as the involvement of an extensive population and the revelation of hidden patterns that are easily overlooked otherwise. The use of social media data in topics like the public opinions towards COVID-19 vaccines has also been reported. For instance, in a recent study based on millions of Twitter data,~\citet{lyu2020social} found that safety, effectiveness, and political issues are chief concerns of the U.S. public when talking about COVID-19 vaccines. However, it is unclear whether the findings revealed by this study are valid with other online communities, which differ from Twitter in user bases and platform features. To better characterize the online discourse related to COVID-19 vaccines, we think it is important and necessary to extend the “ground truth” of public opinions to a broader range of online communities, such as Reddit.

Ranked as the $7^{th}$ most visited website in the United States by November 2020\footnote{https://www.alexa.com/siteinfo/reddit.com} , Reddit is one of the more news-oriented social media platforms according to a research published by the Pew Research Center\footnote{https://www.journalism.org/2016/02/25/seven-in-ten-reddit-users-get-news-on-the-site/}. Unlike Twitter which requests users to fill in profile information, Reddit only asks users to enter usernames in the sign-up process, which results in an anonymous environment. Because of this anonymity feature, users of Reddit, who are referred to as Redditors, perceive a freedom of expression and tend to share their true opinions on topics that they would not openly address on other social media platforms~\cite{daniellereddit}. Therefore, in this work, we collect data from Reddit to examine the thematic characteristics of the discussions of COVID-19 vaccines on it. In doing so, we hope to answer two research questions:

\begin{enumerate}
\item \textbf{RQ 1}: What specific topics characterize the discussions of COVID-19 vaccines on Reddit? How do the topics vary across time and subreddits?
\item \textbf{RQ 2}: Given three most active subreddits, what can we learn from their overlapping users? How do they contribute to the thematic, linguistic similarities and differences among these subreddits? 
\end{enumerate}

Our approach consists of the Latent Dirichlet Allocation (LDA) topic modeling~\cite{blei2003latent}, the Linguistic Inquiry and Word Count (LIWC) text analysis~\cite{pennebaker2001linguistic}, and visual comparisons. Taking into account the nature of social media data, whose development is free from the intervention of researchers, we apply the longitudinal study design to detect the changes in side-wide discussion topics on Reddit over a 9-month time period, and use the cross-section study design to analyze thematic, linguistic similarity, difference, and membership in three different subreddits. Both the longitudinal and cross-section studies are observational studies, which are widely-used approaches in social science fields. Through a combination of computational and qualitative methods, we aim to generate a more comprehensive understanding of the public’s concerns about COVID-19 vaccines, thus paving the way for effective communication strategies geared towards the needs of different communities.

\section{Related Work}
Factors that influence vaccines acceptance and coverage are complex. In the review of articles published between January 2007 and November 2012~\cite{LARSON20142150}, researchers examined 1,164 articles that studied vaccine hesitancy across different regions in the world, concluding that even given a definite number of determinants (i.e., level of income, education, etc.), their impacts on people’s perceptions of vaccines vary across time, place and vaccine types. The study particularly pointed out that as the access to healthcare becomes less a barrier in many countries, personal attitudes and beliefs put an increasing impact on vaccine behaviors, especially in regions wherein vaccine services are available to the most of the population. This finding coheres with what was suggested by another study, which forecasts the trends in vaccination coverage using a time-series analysis over 30 years~\cite{DEFIGUEIREDO2016e726}. Hence, from either a cross-section or a longitudinal perspective, probing into the role of personal opinions in vaccination behaviors at different settings becomes a necessity in the pursuit of a deeper understanding about factors that influence vaccines acceptance.

Entering the $21^{st}$ century, people gradually adapt to the idea that there is no better place than the Internet to find and share information they need at fingertips, including health information. According to a study conducted by the KRC Research\footnote{https://www.webershandwick.com/wp-content/uploads/2018/11/Healthcare-Info-Search-Report.pdf}, 81\% Americans seek for healthcare information online in 2018, with some indicating that they rely more on the Internet than physicians in search of health advice. In the context of vaccine information acquisition and sharing, the overdependence on online information is alarming. When examining online anti-vaccine websites, Kata~\cite{KATA20123778} found that a combination of tactics like shifting hypotheses and censorship makes anti-vaccine claims highly convincing, thus effectively spreading fear and suspicion towards vaccines among Internet users. As social media platforms become prevalent, more and more researchers utilize user-generated contents to understand the public opinions on vaccines as well as their associations with vaccine coverage. For example, with the use of Twitter data,~\citet{DUNN20173033} found a high correlation between the social media information exposure and the state-level HPV vaccine coverage in the U.S., which can explain the differences in coverage that are unexplainable by socioeconomic factors like education and insurance. Further, in a recent study also focusing on the HPV vaccine,~\citet{LamaYuki2019} revealed that political debates are the most frequent topics in all discussions related to HPV vaccines on Reddit, in comparison with conversations on general mainstream media which usually link HPV vaccines to sexual activities. In terms of validity, a study which applied natural language processing methods to Twitter data proved that in the case of understanding vaccine refusal, “the strengths of social media data may greatly outweigh their weakness”~\cite{DREDZE2016550}, which aligns with the consensus of top methodologists that the most reliable research comprises mixed methods and data sources~\cite{AdamsAS1999}. 

Since the outbreak of the COVID-19 pandemic, several studies have already conducted surveys to reflect the public health concerns regarding COVID-19 vaccines (e.g., \citet{pogue2020influences}). Nevertheless, none of the aforementioned studies analyzed Reddit discourse about COVID-19 vaccines and the evolution of underlying concerns implied by the change of topics on this issue. Therefore, we believe our interdisciplinary approach with the use of Reddit data would be a potent complement to this study area. To the best of our knowledge, there is no existing published study conducted under similar conditions. We believe our findings could fill in the gaps currently existing in the understanding of different communities' concerns regarding COVID-19 vaccines, thus offering some useful insights for the design of communication and immunization strategies.

\section{Materials and Methods}
Reddit consists of individual communities or subgroups that differ in topics (i.e.,\textit{ r/Coronavirus}, \textit{r/worldnews}, etc.), which are called subreddits. Reddit users can create original posts, which are referred to as submissions, on a particular subreddit, and comment under submissions. If any user replies to a comment and carries on the conversation under that comment, all the comments together form a comment forest, where it is common that the topics of a comment forest deviates from the original topic discussed in the submission or the top comment. Compared to submissions, which are mostly news, questions, and lengthy stories, comments under submissions are better sources in the light of discourse characterizing, so we decide to focus on top comments under the submissions about COVID-19 vaccines.

To collect the comments, we employ a Reddit API Wrapper called PRAW\footnote{https://praw.readthedocs.io/en/latest/}, which is a Python package extensively used in Reddit-related studies (e.g., Buntain and Golbeck~\cite{buntain2014identifying}). We crawl the comments through the keyword-searching function in the package, so the comments collected must contain at least two keywords: one from the list [`vaccine', `vaccines', `vaccinated', `vaccination', `vacinne', `vacine'] plus one from the list [`covid', `covid-19', `coronavirus', `pandemic', `immunization']. With the combination of these two keyword lists, we are able to crawl only comments  regarding COVID-19 vaccines while excluding those that are related to either other vaccines (i.e., HPV vaccines, flu shots, etc.) or vaccine-unrelated COVID-19 topics. Unlike some studies that collected data from one or two particular subreddits (e.g, \citet{gozzi2020collective}), we crawl comments that meet the conditions in a sitewide basis. For each comment, we download its subreddit, comment id, comment created time, comment author, and comment body. All of the data are visible and accessible to the public. Taking into account the fact that the COVID-19 pandemic did not expand to a world-wide scale until March — the U.S. reported its first coronavirus death on February 29, 2020\footnote{https://www.cdc.gov/media/releases/2020/s0229-COVID-19-first-death.html}, and the WHO declared the pandemic on March 11, 2020 — we discarded comments which predated March 1, 2020. Duplicate comments posted in the same or different subreddits and bots are also pruned. As a result, we had 172,091 comments from 6,466 subreddits that were generated by 107,522 unique users, spanning from March 1, 2020 to December 15, 2020. Table~\ref{tab:top_8_subreddits} lists the top eight subreddits ranked by the number of comments, along with their general attributes. The total numbers of comments and authors of these 8 subreddits constitute over 1/3 of the entire comments and authors in all the subreddits.

\begin{table*}[htbp!]
\renewcommand{\arraystretch}{1.2}
\scriptsize
\centering
\caption{Top eight subreddits with the most number of comments.}

\begin{tabular}{@{}ccccccc@{}}

\hline
\textbf{Subreddits} & \textbf{Num. of Comments} & \textbf{Num. of Authors} & \textbf{Average Comments per Author (Std)} & \textbf{Median}
& \textbf{Max}   &   \textbf{Min}\\
\hline
r/Coronavirus         & 17,263                    & 9,849                & 1.75 (3.18)  &  1    &  114   &  1                             \\
r/worldnews           & 7,985                     & 5,782                & 1.38 (8.93)   &  1    &   676    &  1                        \\
r/conspiracy          & 7,122                     & 4,159                & 1.71 (3.02)   &  1    &   145   &  1                         \\
r/politics            & 7,006                     & 5,520                & 1.27 (1.55)   &  1     &    71   &   1                             \\
r/wallstreetbets      & 6,424                     & 3,982                & 1.61 (2.14)   &  1     &    58   &   1                        \\
r/AskReddit           & 5,700                     & 4,954                & 1.15 (1.10)   &  1     &    55    &  1                          \\  
r/news                & 3,779                     & 3,123                & 1.21 (0.79)   &  1     &    20   &  1                       \\
r/COVID19             & 2,378                     & 1,169                & 2.03 (4.12)   &  1     &   64   &  1                         \\
\hline
\textbf{Grand Total}         & \textbf{57,657}                    & \textbf{35,798 }            & \textbf{1.61 (4.78)}                &   \textbf{1}     &  \textbf{747}     &    \textbf{1}                     \\
\hline
\end{tabular}

\label{tab:top_8_subreddits}
\end{table*}

\subsection{LDA}
Based on the evidence presented in a recent study~\cite{albalawi2020using}, LDA is not only the most popular topic modeling method to date but also achieves better performance in a short-text context among many other topic modeling methods such as the Latent Semantic Analysis (LSA). Moreover, the study also proves that LDA can produce higher-quality and more coherent topics than other unsupervised topic modeling algorithms. Therefore, LDA is chosen for topics extraction in our study. Before feeding the comments into the LDA model, we remove punctuations from the comments and conduct lemmatization. We also process the data by using the directory of ``english'' stopwords from the Natural Language Toolkit (NLTK)\footnote{https://www.nltk.org/book/ch02.html} with an extended list of words [`vaccine', `vaccines', `vaccinated', `covid', `coronavirus', `virus', `us', `get', `take', `also'], which aims to best differentiate possible topics. To determine the most appropriate number of topics, we train LDA models with different numbers of topics using Gensim\footnote{https://radimrehurek.com/gensim/} and graph their respective coherence scores. The highest coherence score (0.540) is reached when the number of topics is set to 38. However, after manually inspecting the keywords of the 38 topics, we find that many topics are overlapping with each other. Therefore, we make several attempts by changing the number of topics until a number is found that both produces coherent topics and has a relatively high coherence score, which occurs at 10. The coherence score of the chosen model is 0.511.

\subsection{LIWC}
We apply LIWC 2015 to capture the psychological information from the collected comments. It reads a given text and counts the percentage of words that reflect different emotions, thinking styles, social concerns.\footnote{https://liwc.wpengine.com/how-it-works/} \citet{lyu2020social} have suggested that sentiment is positively related to the acceptance level of COVID-19 vaccines, therefore, {\tt posemo} (i.e., positive emotion) and {\tt negemo} (i.e., negative emotion) categories are selected to measure the sentiment expressed through the comments. To get more nuanced insights into emotions, {\tt anx} (i.e., anxiety) and {\tt anger} are included. {\tt health}, {\tt risk}, {\tt death} are also included to obtain a broader understanding of the comments. Similar to the methods of~\citet{chen2020eyes}, we concatenate all the comments of the authors of the same group and apply LIWC 2015. Since the scales of the LIWC categories are not the same, we normalize the LIWC scores in the plots to better illustrate the the qualitative differences.

\section{Results}

\subsection{RQ 1 Results}

Table~\ref{tab:top10_topic} itemizes the 10 topics that predominate the comments. According to the keywords under each topic and their probability of occurring in the corresponding topic, we manually designate topic labels to the topics generated by the model, which is a common practice in studies that also use LDA for topic modeling (e.g., \citet{okon2020natural}). Note that every comment may belong to more than one topic, while in this case we categorize each comment to only its dominant topic in order to calculate the proportion of comments for each topic. The first row indicates that among all the comments, 12.44\% of them belong to the topic ``skeptical/aggressive remarks", and the top 10 keywords associated with this topic are listed in a descending order of contribution. Not only does the first topic occupy the highest percent of comments, but the percentage difference between it and the second topic ``clinical trials/research/testing" is larger (1.22\%) than the difference between any other two consecutive topics. The Appendix lists the representative example of comments for each topic, respectively.

\begin{table*}[htbp!]
\scriptsize
    \setlength{\tabcolsep}{1pt}
    \renewcommand{\arraystretch}{1.2}
    \centering
    \caption{10 topics generated by LDA and their associated keywords.}
    \begin{tabular}{@{}ccc@{}}
    \hline
    \textbf{Topics} & 
    \textbf{\% of Comments} & 
    \textbf{Keywords} \\
    \hline
    skeptical/aggressive remarks     &	  
    12.44\% & 
    people, mask, wear, make, fuck, anti, thing, literally, shit, stop \\
    \hline
    clinical trials/research/testing   &
    11.22\% &
    test, trial, study, phase, result, dose, effective, testing, early, datum \\
    \hline
    life/family/kids   &
    11.18\% &
    work, live, home, feel, day, life, school, family, child, back \\
    \hline
    people/vaccine efficacy/risks   &
    10.35\%  &
    people, immunity, death, risk, die, rate, population, case, high, 
    number \\
    \hline
    governments/big companies   &
    10.16\%   &
    government, make, company, world, country, pandemic, work, money, pay, develop  \\
    \hline
    symptoms/immune systems   &
    9.38\%    &
    effect, immune, antibody, response, human, system, develop, body, side, cell \\
    \hline
    time/long-term effects    &
    9.15\%    &
    year, long, time, month, thing, good, make, term, happen, bad \\
    \hline
    stock market/sports      &
    8.96\%     &
    back, market, start, big, year, stock, business, buy, play, hold   \\
    \hline
    politics/news sources     &
    8.80\%    &
    trump, https, comment, question, article, post, news, source, claim, pandemic    \\
    \hline
    lockdown/spread/cases     &
    8.36\%    &
    case, lockdown, health, country, spread, hospital, social, state, open, place \\
    \hline
    \end{tabular}
    
    \label{tab:top10_topic}
\end{table*}

\begin{figure*}[htbp!]
\centering
\includegraphics[width=0.8\linewidth]{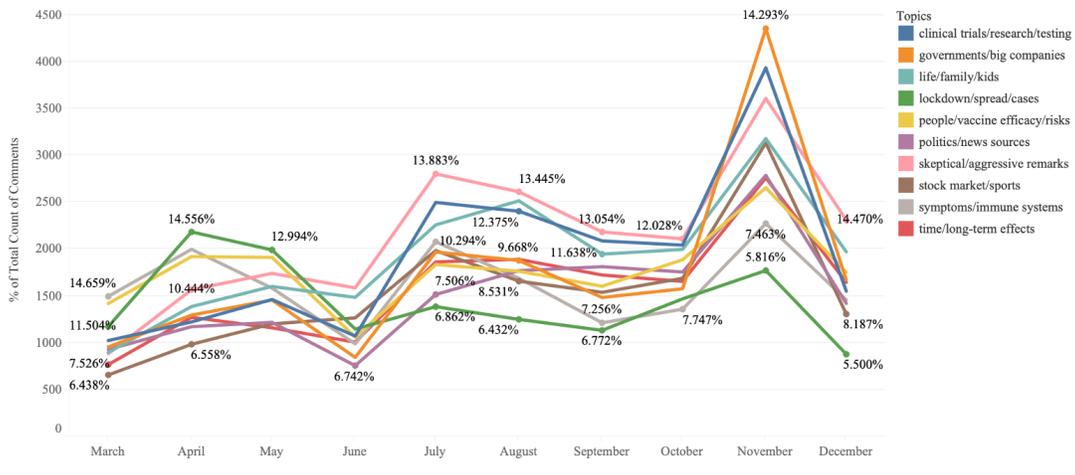} 
\caption{Trends of the proportions of comments of different topics from March 1, 2020 to December 15, 2020.}
\label{fig:trends_topic}
\end{figure*}

As shown in the example of ``skeptical/aggressive remarks", one common feature of this type of comments is that they convey conspicuous suspicions or condemnations of certain entities, which could be anti-vaxxers, vaccine manufacturers, governments, etc. From a longitudinal perspective, Figure~\ref{fig:trends_topic} manifests that comments of ``skeptical/aggressive remarks" predominated the overall discourse related to COVID-19 vaccines from June to December except for November (6 out of 10 months). Even though comments of ``governments/big companies" supplanted the dominant position of comments belonging to ``skeptical/aggressive remarks" in November, the number of the latter still outstripped the number of comments of any other topics till the end of the collection period. In contrast, comments about ``lockdown/spread/cases" originally had the largest proportion among all comments in April and May, but the comments' proportion decreased sharply from May to June, and remained as the least discussed topic till the end. In a holistic view, there is a salient spike caused by the increase of the amount of all comments in November, which may be attributed to real-world events carrying prominent importance, such as the 2020 U.S. Presidential Election and BioNTech and Pfizer's announcement that their vaccine is 95\% effective.  

\begin{figure*}[htbp!]
\centering
\includegraphics[width=0.8\linewidth]{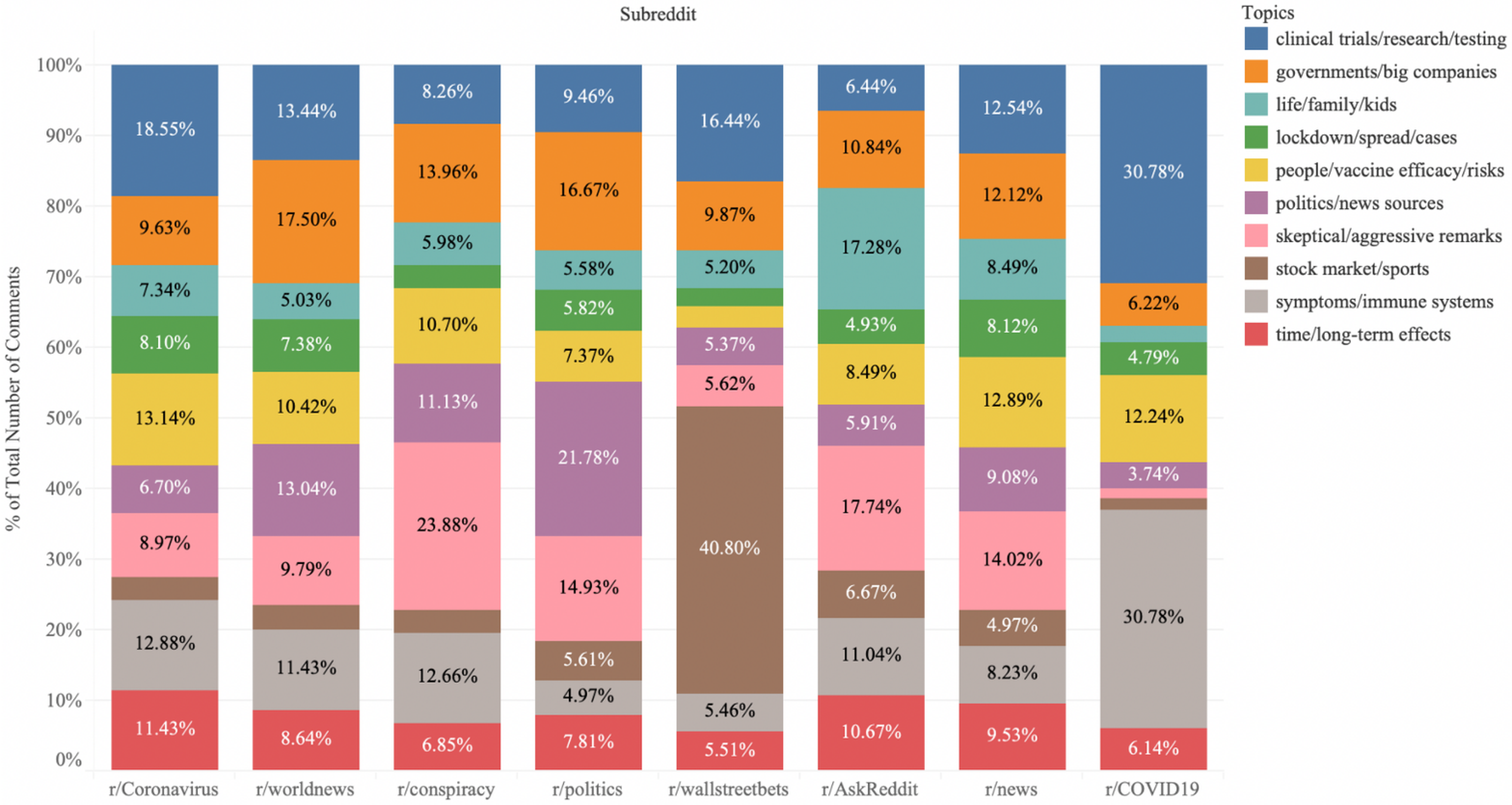} 
\caption{Distributions of comments of different topics across top eight subreddits.}
\label{fig:topics_of_8_subreddits}
\end{figure*}

From a cross-section perspective, Figure~\ref{fig:topics_of_8_subreddits} illustrates the proportions of comments of different topics in the top eight subreddits. Particularly, comments of ``skeptical/aggressive remarks" occupy the largest proportion in \textit{r/conspiracy}; comments of ``stock market/sports" have the largest proportion in \textit{r/wallstreetbets}; comments of ``clinical trials/research/testing" and ``symptoms/immune systems" predominate \textit{r/COVID19}. It is noteworthy that the dominant topic detected by our LDA model in each subreddit is in accordance with the subreddit's description. For example, in the ``About Community" field of \textit{r/COVID19}, it reads ``In December 2019, SARS-CoV-2, the virus causing the disease COVID-19, emerged in the city of Wuhan, China. This subreddit seeks to facilitate scientific discussion of this global public health threat"\footnote{https://www.reddit.com/r/COVID19/}. This finding keeps constant in the rest of the subreddits, which proves the robustness of our topic characterizing model. Moreover, the proportion of comments of the same topic varies widely from subreddit to subreddit, which may be caused by two conditions: Reddit users tend to discuss particular topics in particular subreddits, or the fundamental user bases of different subreddits differ and have disparate concerns. In the first case, it is difficult for us to prove whether the comments made by users can represent their major concerns, since the variation may also be caused by the policy of a subreddit; while in the second case, we can prove that the comments made in a specific subreddit can characterize primary concerns of its users, as long as its user base hardly overlaps with that of other subreddits.

\subsection{RQ 2 Results}
To understand whether the disparity in proportions of comments derive from the differences in subreddits' user bases, we pick three subreddits out of the top eight subreddits: \textit{r/Coronavirus}, \textit{r/worldnews}, and \textit{r/conspiracy}. They are the subreddits that have the most comments among the collected data. In Table~\ref{tab:label_standard}, we divide them into three pairs and compare each pair's overlapping user bases, their most discussed topics in each subreddit as well as their linguistic profiles. 

\begin{table*}[htbp!]
\scriptsize
    \centering
    \setlength{\tabcolsep}{3pt}
    \renewcommand{\arraystretch}{1.2}
     \caption{Characteristics of comment authors who participated in multiple subreddits and most commonly discussed topics by these overlapping authors.}
    \begin{tabular}{|c|c|c|}
    \hline
    \textbf{} &  
  \multicolumn{1}{|c|}{\textbf{r/Coronavirus}}  &
  \multicolumn{1}{|c|}{\textbf{r/worldnews}} \\
    \hline
    \textbf{Num. of Overlapping Comment Authors } & 435 & 435    \\
    \hline
    \textbf{Num. of Total Comment Authors}	& 1682 & 717 \\
    \hline
    \textbf{Mean of Num. of Comments} & 3.87** & 1.65** \\
    \hline
    \multirow{5}{15em}{\textbf{Top Five Occurring Topics (\% of the overall comments)}} & 
    clinical trials/research/testing (26.46\%)** 
    & clinical trials/research/testing (20.22\%)** \\
    & symptoms/immune systems (15.10\%)
    & governments/big companies (17.57\%)\footnote[2]{} \\
    & governments/big companies (14.80\%)\footnote[2]{}
    & symptoms/immune systems (16.04\%) \\
    & people/vaccine efficacy/risks (12.25\%)
    & people/vaccine efficacy/risks (12.41\%) \\
    & time/long-term effects (9.39\%)
    & lockdown/spread/cases (8.09\%)\footnote[2]{} \\
    \hline
    \textbf{} &  
    \multicolumn{1}{|c|}{\textbf{r/Coronavirus}}  &
    \multicolumn{1}{|c|}{\textbf{r/conspiracy}} \\
    \hline
    \textbf{Num. of Overlapping Comment Authors } & 104 & 104    \\
    \hline
    \textbf{Num. of Total Comment Authors}	& 183 & 220 \\
    \hline
    \textbf{Mean of Num. of Comments} & 1.76 & 2.12 \\
    \hline
    \multirow{5}{15em}{\textbf{Top Five Occurring Topics (\% of the overall comments)}} & 
    people/vaccine efficacy/risks (17.49\%) 
    & people/vaccine efficacy/risks (15.45\%) \\
    & governments/big companies (14.75\%)
    & governments/big companies (15.45\%) \\
    & symptoms/immune systems (13.66\%)
    & symptoms/immune systems (14.55\%) \\
    & clinical trials/research/testing (13.11\%)
    & skeptical/aggressive remarks (14.09\%)\footnote[2]{} \\
    & time/long-term effects (11.48\%)\footnote[2]{}
    & politics/news sources (10.91\%) \\
    \hline
    \textbf{} &  
    \multicolumn{1}{|c|}{\textbf{r/conspiracy}}  &
    \multicolumn{1}{|c|}{\textbf{r/worldnews}} \\
    \hline
    \textbf{Num. of Overlapping Comment Authors } & 91 & 91    \\
    \hline
    \textbf{Num. of Total Comment Authors}	& 192 & 129 \\
    \hline
    \textbf{Mean of Num. of Comments} & 2.11** & 1.42** \\
    \hline
    \multirow{5}{15em}{\textbf{Top Five Occurring Topics (\% of the overall comments)}} & 
    skeptical/aggressive remarks (16.67\%) 
    & clinical trials/research/testing (14.73\%) \\
    & people/vaccine efficacy/risks (15.10\%)
    & people/vaccine efficacy/risks (14.73\%) \\
    & symptoms/immune systems (14.58\%)
    & symptoms/immune systems (13.95\%) \\
    & clinical trials/research/testing (13.02\%)
    & skeptical/aggressive remarks (13.95\%) \\
    & governments/big companies (12.50\%)
    & governments/big companies (13.17\%) \\
    \hline
    \multicolumn{3}{l}{\textit{Note.} \footnote[2]{} p-value $<$ 0.1; *\footnotesize{p-value $<$ 0.05}; **\footnotesize{p-value $<$ 0.01}} \\
    \end{tabular}

    \label{tab:label_standard}
\end{table*}

\subsection{r/Coronavirus vs. r/worldnews}
\begin{figure}[htbp!]
    \centering
    \includegraphics[width = 0.47\textwidth]{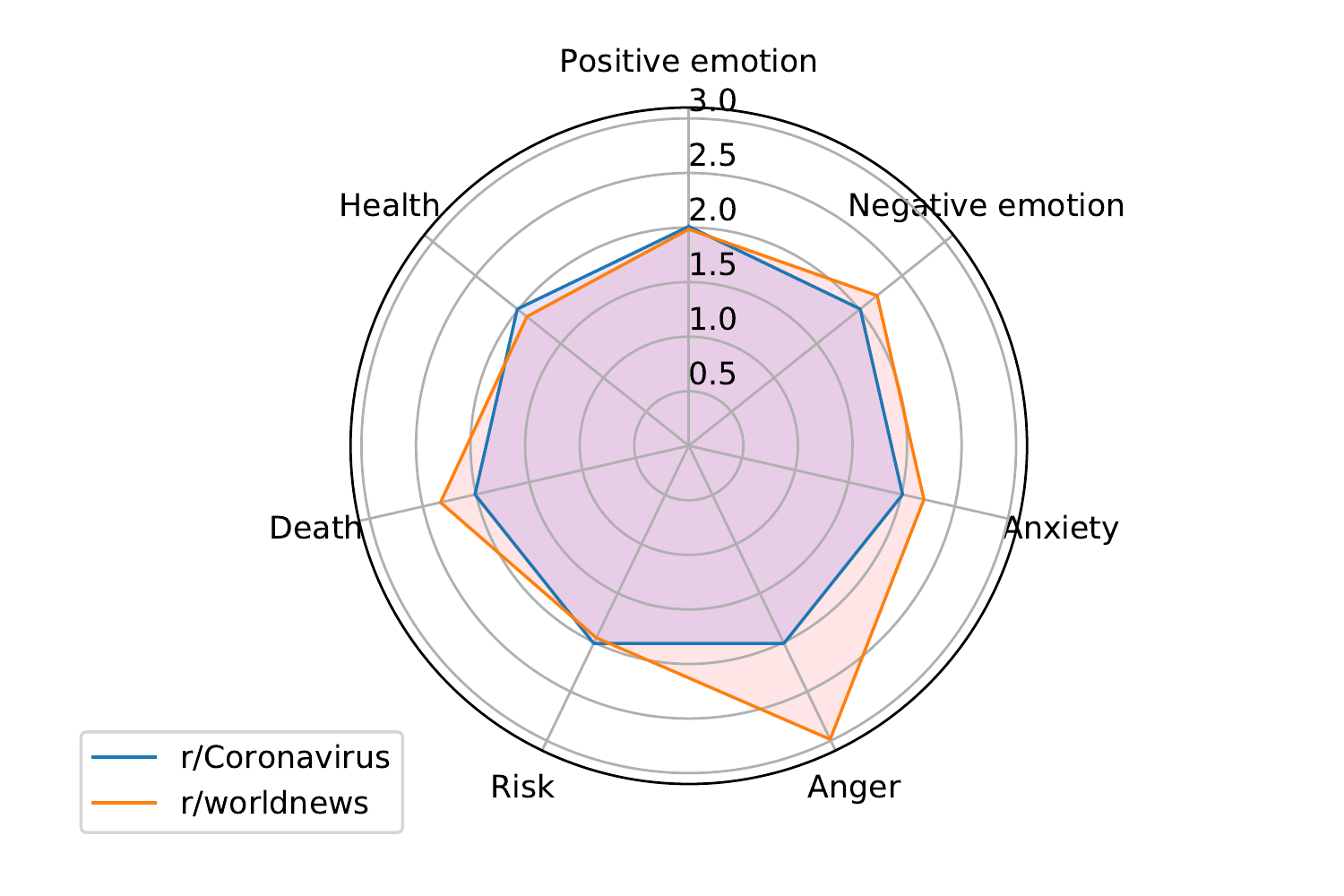}
    \caption{Linguistic profiles for the comments of the overlapping users of \textit{r/Coronavirus} and \textit{r/worldnews}. (Note: The scores have been normalized.)}
    \label{fig:covid_world}
\end{figure}

This pair shares the most overlapping users (435) among all three pairs. Respectively, the proportions of overlapping users in the user bases of these two subreddits are 4.42\% (435 out of 9,849) and 7.52\% (435 out of 5,782). In average, each of the 435 users posted four comments in \textit{r/Coronavirus} and two comments in \textit{r/worldnews}, which implies that these overlapping users are more active in \textit{r/Coronavirus} than \textit{r/worldnews} during the selected time period. In both subreddits, the topic discussed most by these users is ``clinical trials/research/testing", whereas the proportion of comments related to this topic is relatively larger in \textit{r/Coronavirus} than in \textit{r/worldnews}, which echoes the findings in the linguistic profiles where the overlapping users pay less attention to the health-related issues in \textit{r/worldnews}. Conversely, the proportions of comments related to ``governments/big companies'' and ``lockdown/spread/cases'' are both higher in \textit{r/worldnews}. Interestingly, as shown in Figure~\ref{fig:covid_world}, the overlapping users express more negative emotions, anxiety, anger, and show more concerns about death-related issues in \textit{r/worldnews} than \textit{r/Coronavirus}.

\subsection{r/Coronavirus vs. r/conspiracy}
\begin{figure}[htbp!]
    \centering
    \includegraphics[width = 0.47\textwidth]{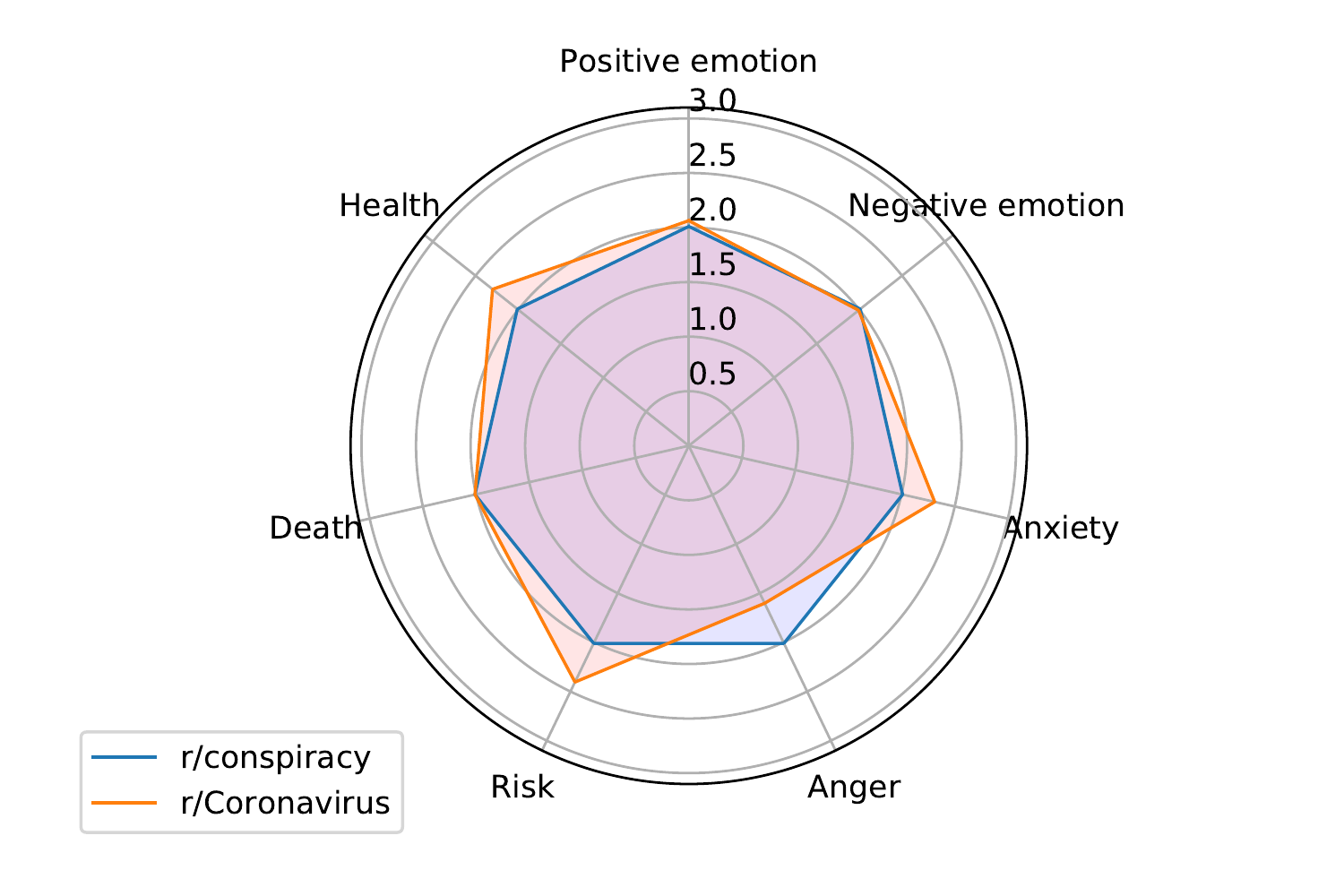}
    \caption{Linguistic profiles for the comments of the overlapping users of \textit{r/Coronavirus} and \textit{r/conspiracy}. (Note: The scores have been normalized.)}
    \label{fig:covid_conspiracy}
\end{figure}

This pair shares 104 overlapping users, each of whom posted an average of 2 comments in both subreddits during the selected period of time. Respectively, the overlapping users occupy 1.06\% (104 out of 9,849) and 2.50\% (104 out of 4,159) of the overall user bases of \textit{r/Coronavirus} and \textit{r/conspiracy}. The top five topics that characterize the comments of the overlapping users in these two subreddits are nearly the same, with slight variations in the proportions of comments of each topic. In particular, the overlapping users talked more about ``time/long-term effects" in \textit{r/Coronavirus}, while they made more comments about ``skeptical/aggressive remarks" in \textit{r/conspiracy}. The overlapping users show more anxiety, pay more attention to risks, and talk more about health-related issues in \textit{r/Coronavirus}, which corresponds to the higher proportion of the comments about ``time/long-term effects". In addition, more anger is observed among the overlapping users in \textit{r/conspiracy} who talked more about ``skeptical/aggressive remarks" (Figure~\ref{fig:covid_conspiracy}).

\subsection{r/conspiracy vs. r/worldnews}
\begin{figure}[htbp!]
    \centering
    \includegraphics[width = 0.47\textwidth]{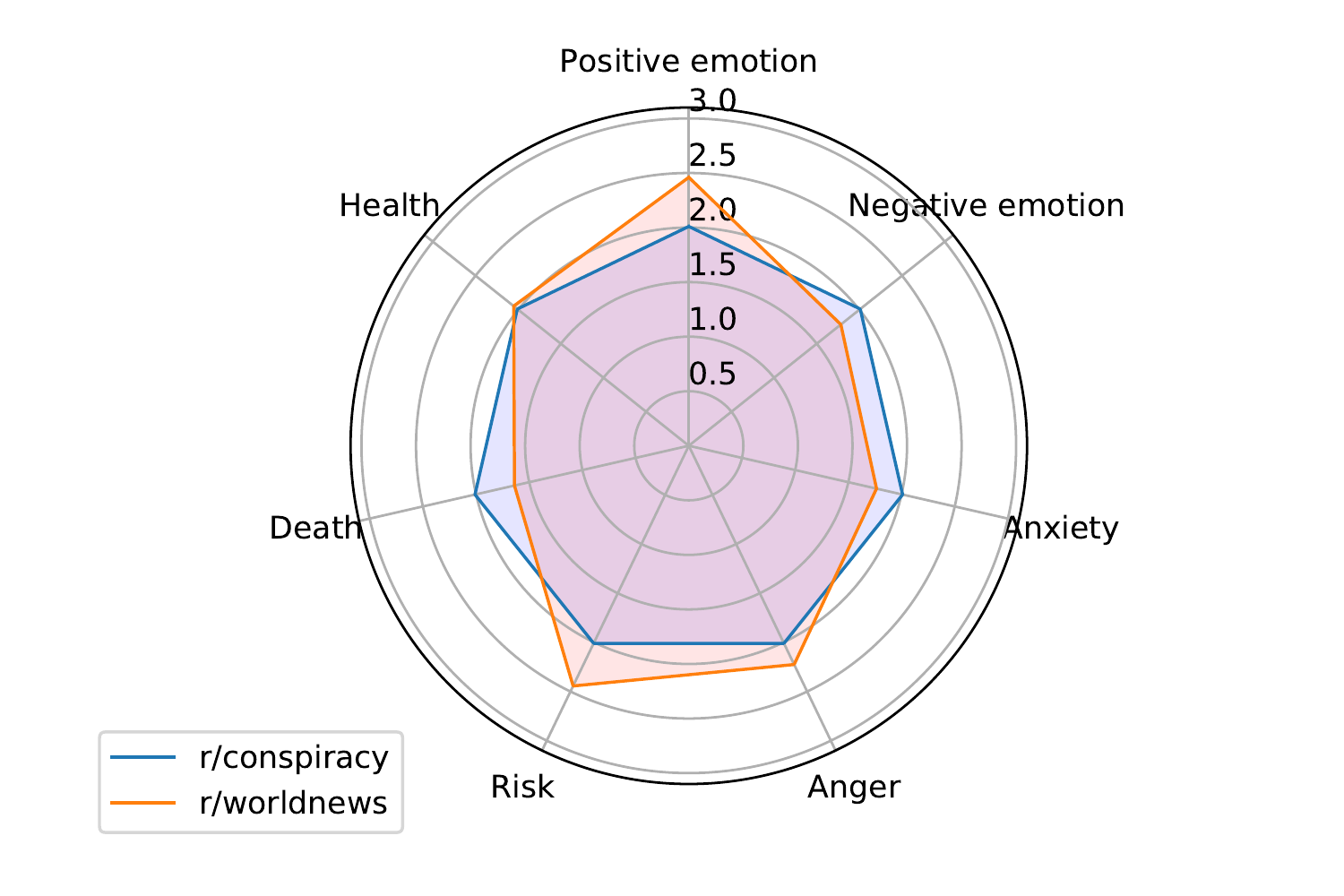}
    \caption{Linguistic profiles for the comments of the overlapping users of \textit{r/conspiracy} and \textit{r/worldnews}. (Note: The scores have been normalized.)}
    \label{fig:conspiracy_world}
\end{figure}

This pair has the least overlapping users (91) among the three pairs. The proportions of overlapping users in the user bases of these two subreddits are 2.19\% (91 out of 4,159) and 1.57\% (91 out of 5,782). These users made an average of two comments in \textit{r/conspiracy} and and an average of one comment in \textit{r/worldnews}. Since the difference between the means is statistically significant, it is reasonable to declare that these overlapping users are more active in \textit{r/conspiracy} than \textit{r/worldnews}. In terms of the top five occurring topics, although comments related to ``skeptical/aggressive remarks" seem to be more frequent in \textit{r/conspiracy}, the z-test shows no statistically significant difference between the two subreddits' proportions of comments related to any of the listed topics. As for the linguistic profiles, the overlapping users express more positive emotions, focus more on risks, and show more anger in \textit{r/worldnews} (Figure~\ref{fig:conspiracy_world}). 

\par
\bigskip
In general, \textit{r/conspiracy} shares a relatively small number of overlapping users with either \textit{r/Coronavirus} or \textit{r/worldnews}. Despite some nuances in the proportions of different types of comments, the main topics discussed by overlapping users in disparate subreddits are nearly identical, which refutes the hypothesis that the same users discuss different topics based on the features/policies of specific subreddits and thus contribute to the variations in different comments' proportions to a large extent. In other words, the impact of the overlapping users is modest. Therefore, it is valid to infer that the variations in topic proportions across subreddits are mainly attributed to different user bases, which is confirmed by the small percentage of overlapping users in the overall user base of each subreddit. A further qualitative analysis of the linguistic profiles of the comments of overlapping users is conducted. The differences in the linguistic profiles, combined with topics have revealed more insights into the user bases. It is especially worth noting that \textit{r/conspiracy} exceeds \textit{r/Coronavirus} in anger and \textit{r/worldnews} in death, anxiety, and negative emotion, respectively. Instead of what the word ``conspiracy'' suggests about the subreddit, we find that comments from this subreddit vary in stances on COVID-19 vaccines: some of them question the intents of governments/big companies with suspicions and malignity, while others condemn those who are suspected as anti-vaxxers. One common linguistic feature of these comments is the dense use of offensive/profane words, which also contributes to the dominant proportion of ``skeptical/aggressive remarks" in the subreddit. To reveal the complexity of opinions, we include representative examples from \textit{r/conspiracy} in the Appendix.

\section{Discussion}
By examining what specific topics characterize the discussions of COVID-19 vaccine on Reddit, we find that the proportion of the comments about ``skeptical/aggressive remarks'' outweighed that of other topic during the pandemic. Moreover, these topics vary across time and subreddits, suggesting differences in subreddits' user bases. We further pick three subreddits out of the top eight: \textit{r/Coronavirus}, \textit{r/worldnews}, \textit{r/conspiracy}, and find that the variations in topic proportions across subreddits are mainly related to the thematic and linguistic differences among the overlapping users. 

\subsection{Implications}
As the first work to characterize discourse related to COVID-19 vaccines on Reddit, our study has three significant implications. First and foremost, aside from Twitter~\cite{jamison2020not}, Reddit has served as a ``hotbed’’ for conspiracy theories and disinformation since the outbreak of the pandemic. The facts that \textit{r/conspiracy} has gained the third most comments and that the sitewide conversations have been dominated by those belonging to ``skeptical/aggressive remarks” for six months are simply shocking. Nevertheless, this finding is in accordance with the claim that the $21^{st}$ century is ``the golden age of anti-vaccine conspiracies”, made by Stein~\cite{stein2017golden}. Second, despite the small number of overlapping users between subreddits, each of the most active subreddits has its own user bases. Hence, reliable news posted in \textit{r/Coronavirus} would not draw attention from users in \textit{r/conspiracy}, while the latter may be the group of people who need reliable information sources the most. Lastly, even on a single social media platform like Reddit, the concerns related to COVID-19 vaccines vary from subcommunity to subcommunity, not to mention more popular social media platforms like Facebook and Twitter. These various concerns imply different communication preferences~\cite{gilkey2017parents}. Therefore, in real practice, we must consider the characteristics of different groups and implement communication strategies targeting at specific groups’ needs. \textit{Speak the same language} is the weapon for us to overcome the overabundance of mis- and disinformation in times of crisis. 
\par
In terms of tackling the challenges suggested by the findings, active actions and collaborations from multiple parties are needed. As stated by policy advisors Wardle and Singerman, ``we need responses that acknowledge the complexity of defining misinformation, of relying on scientific consensus, and of acknowledging the power of narratives"~\cite{wardle2021too}. Our study contributes to the first stage by quantifying multifaceted discussions regarding COVID-19 vaccines and further disentangle them using NLP techniques. Our study also sheds light on the disagreements among different subgroups by providing concrete examples and comparing them on a user- as well as thematic-level. Based on what we find and other pertinent studies, social media platforms and policy makers should obtain a more comprehensive landscape of online discussions and thus be able to create more effective strategies that build the public's consensus and trust in vaccinations. 

\subsection{Limitations}
Admittedly, our study is not free of limitations. Although our model achieves a high coherence score in characterizing the collected comments, the generated topics can only represent the concerns of users who posted top comments that meet our selection standards (keyword searching, etc.). There may be some opinions or concerns underlying comments that were not taken into account in our current analysis, so in order to obtain a more precise picture of thematic characteristics of the discourse related to COVID-19 vaccines on Reddit, we should amplify the scale of data in future works. In addition, since it is impossible to know Reddit users’ demographics, our findings simply imply the discourse characteristics of the general Reddit users instead of the users from a certain region or socioeconomic level. From the perspective of designing communication or immunization programs, more information is needed to target the needs of specific groups. Moreover, we did not incorporate the interactions between Reddit users (i.e., upvotes and downvotes) in this work, although these are elements that may help us better understand the dynamics of pertaining conversations--- this can be done in future works.

\section{Conclusion}
In this study, we characterize the Reddit discourse related to COVID-19 vaccines through a combination of computational and qualitative methods. Specifically, we employ an LDA model to categorize top Reddit comments from March 1, 2020 to December 15, 2020 under 10 topics, analyze how the number of each type of comments changed over time, and examine these comments’ proportions across eight different subreddits--- we detect discrepancies from both longitudinal and cross-section perspectives. Thereafter, we conduct a more careful investigation into the thematic and linguistic variations between subreddits by comparing three most active subreddits. With the use of statistical tests, we confirm that although there are overlapping users between these subreddits, the scale of them is quite smaller than that of each subreddit’s own user base, thus proving the hypothesis that thematic variations among subreddits’ conversations mainly result from differences in user bases. This finding also suggests the need of taking care of different subgroups’ communication preferences within a large social media platform. Furthermore, we employ LIWC and find that there are also differences among the overlapping users across emotions, personal concerns and attention. As information on social media platforms puts greater influence on people’s vaccination behaviors~\cite{KATA20123778}, we believe the findings of our study can provide policy makers with useful insights in the process of designing effective communication and immunization programs.

\section*{Acknowledgments}


\subsection*{Author Contributions} 
Describe contributions of each author to the paper, using the first initial and full last name. 
W. Wu, H. Lyu and J. Luo conceived the idea and designed the experiments. W. Wu and H. Lyu conducted the experiments. W. Wu wrote the majority of manuscript. H. Lyu wrote part of the manuscript. J. Luo supervised the study and reviewed the manuscript.





\subsection*{Funding}
No external funding was received.




\subsection*{Conflicts of Interest}
The authors declare that there is no conflict of interest regarding the publication of this article.



\subsection*{Data Availability}
Anonymized data created for the study are or will be available in a persistent repository upon publication.

\bibliography{name}
\bibstyle{aaai}

\appendix
\section{Appendix}

\subsection{Example comments under the 10 topics}

\begin{quote}
\textbf{Topic 1: clinical trials/research/testing}
\begin{small}
{\tt \small
``Normally it takes a long time to enrol enough people. It also takes a while for people to become infected with the thing that you're studying. We don't have that problem in a pandemic where almost 5 million people every week is infected with COVID-19. It's the same process, but it doesn't take as long to get results. [...]" 
}
\end{small}

\textbf{Topic 2: governments/big companies}
\begin{small}
{\tt \small
``As someone who is vaccinated and whose children are vaccinated I won't touch this vaccine with a 10 foot pole. When its so safe that the manufacturers demand blanket protection from any Liability from side effects and the countries governments give it to them that should already be a huge red flag. Pfizer: our vaccine is super safe and will help stop Covid-19. Also Pfizer: we're going to need complete immunity from liability if our vaccine turns out to be a complete fiasco. [...]" 
}
\end{small}

\textbf{Topic 3: life/family/kids}
\begin{small}
{\tt \small
``It really concerns me for my children more than myself.  I don't want to give them a chemical during their childhood that may fuck them up medically later in life.  We homeschool them.  We don't go out unnecessarily, we wear masks, we social distance, etc.  I'm not against ever taking a covid vaccine but I will definitely be waiting before taking it or giving it to my kids. [...]" 
}
\end{small}

\textbf{Topic 4: lockdown/spread/cases}
\begin{small}
{\tt \small
``[...] My state — <state name> — has been completely free of any community transmission for over seven months and life has basically returned to normal. The only cases have been people returning from overseas who have to undergo a mandatory quarantine for two weeks when they arrive." 
}
\end{small}

\textbf{Topic 5: people/vaccine efficacy/risks}
\begin{small}
{\tt \small
``Something like 12\% of the population have Hashimoto's. If the condition was a significant risk factor with a vaccine, it WOULD have shown up in the clinical trials. Even under the worst possible assumptions, the risks from Covid would be literally a thousand times higher than from a vaccine. Anyone who needs any convincing for taking the vaccine should visit the Covid long-hauler board." 
}
\end{small}

\textbf{Topic 6: politics/news sources}
\begin{small}
{\tt \small
``Cool. In the meantime you could actually read about vaccine distribution, here's some useful links: <source links>" 
}
\end{small}

\textbf{Topic 7: skeptical/aggressive remarks}
\begin{small}
{\tt \small
``[...] See you out in the real world very very soon vaccine is coming, wooooohoooo! Can’t wait to burn my mask in dumpster fire! Vaccine vaccine vaccine!! Goodbye covid!!!!! Even <name> and <name> are like who gives a fukkkkk!"}
\end{small}

\textbf{Topic 8: stock market/sports}
\begin{small}
{\tt \small
``The fact so many stocks have risen to higher valuations then the Pre-March covid crash is just laughable. All these sad people will be left holding bags once the Rich take a massive dump on all of them. Everything is essentially at an ATH, Over valued and propt up by Vaccine optimism/Governments Stimulus. The rug will be pulled out from under the average persons feet trying to invest for retirement." 
}
\end{small}

\textbf{Topic 9: symptoms/immune systems}
\begin{small}
{\tt \small
``I am O- and have had Covid. I had moderate symptoms for 36-48 hours (low fever, night sweats, chills, body aches, diarrhea). I did not have respiratory issues and did not develop a cough. I’m an extra complicated case, however, as within the past six years, I have underwent 2.5 years of chemo for acute lymphoblastic leukemia (ALL). I am quite terrified of how Covid or the vaccine might affect the ALL, which is currently in remission. Thought I’d throw this out there in case anyone smarter than me has any insights." 
}
\end{small}

\textbf{Topic 10: time/long-term effects}
\begin{small}
{\tt \small
``[...] Like you said, it has been less than a year, we don't known jack shit yet, but they want to inject us all with rushed vaccines we don't even know the effectiveness for or how long that effectiveness will last. And you're okay with that?" 
}
\end{small}

\end{quote}

\subsection{Example comments from \textit{r/conspiracy}:} \bigbreak

{\tt \small``Vaccines wouldn't have been necessary, and this fucking pandemic wouldn't still be such a major issue if this ridiculous administration hadn't dissolved the group specifically designed to combat this process and if the population at large had followed simple fucking suggestions. But they can't and here we are... saying 'No' to a fucking virus."}

{\tt \small ``Somebody who's anti vaxx wouldn't accept a COVID vaccine no matter what the top health officials said.
They lied to us about aids and now they are lying about COVID. Do not take the test do not take the vaccine it’s the aids epidemic all over again!!!!
vaccines are cool. I'm probably been vaccinated more times than anyone on this sub or even the pro-vax subs, and I even got a vaccine for anthrax for no reason once and it hurt like hell. It's like getting hit in the arm by a boxer and doesn't go away for a couple days."}

{\tt \small ``anyway, like I said, vaccines are fine, but in this case they're not. It's clear that there could be a way to treat COVID 19 without a worldwide mandatory vacccine, yet any research regarding that is being heavily suppressed and unfairly criticized by the mainstream media and Dr. Fauci. Does that seem like good science? Not really."}

\end{document}